\documentclass[aps,prb,showpacs,twocolumn,preprintnumbers]{revtex4-1}
\usepackage{mathrsfs}
\usepackage{amsmath}
\usepackage{amsfonts}
\usepackage{amssymb}
\usepackage{amsthm}
\usepackage{graphicx}
\usepackage{natbib}
\usepackage{color}
\usepackage{hyperref}
\usepackage{bm}
\usepackage[caption=false]{subfig}
\usepackage{verbatim}
\providecommand{\U}[1]{\protect\rule{.1in}{.1in}}

\begin{document}
\title{Spin Hall noise}
\author{Akashdeep Kamra$^{1,2}$}
\author{Friedrich P. Witek$^{1}$}
\author{Sibylle Meyer$^{1,4}$}
\author{Hans Huebl$^{1,3}$}
\author{Stephan Gepr{\"a}gs$^{1}$}
\author{Rudolf Gross$^{1,3,4}$}
\author{Gerrit E. W. Bauer$^{5,6,2}$}
\author{Sebastian T. B. Goennenwein$^{1,3}$}
\affiliation{$^1$Walther-Mei{\ss}ner-Institut, Bayerische Akademie der 
Wissenschaften, 85748 Garching, Germany}
\affiliation{$^2$Kavli Institute of NanoScience, Delft University of Technology, 
2628 CJ Delft, The Netherlands}
\affiliation{$^3$Nanosystems Initiative Munich (NIM), 80799 Munich, Germany}
\affiliation{$^4$Physik-Department, Technische Universit{\"a}t M\"unchen, 85748 
Garching, Germany}
\affiliation{$^5$Institute for Materials Research, Tohoku University, Sendai 
980-8577, Japan}
\affiliation{$^6$WPI Advanced Institute for Materials Research, Tohoku University, Sendai 980-8577, Japan}

\begin{abstract}
 We measure the low-frequency thermal fluctuations of pure spin current in a Platinum film deposited on yttrium iron garnet via the inverse spin Hall effect (ISHE)-mediated voltage noise as a function of the angle $\alpha$ between the magnetization and the transport direction. The results are consistent with the fluctuation dissipation theorem in terms of the recently discovered spin Hall magnetoresistance (SMR). We present a microscopic description of the $\alpha$ dependence of the voltage noise in terms of spin current fluctuations and ISHE.
\end{abstract}



\maketitle


\section{Introduction}
The quote ``The noise is the signal'' by Rolf Landauer~\cite{landauer1998} emphasizes the usefulness of noise spectroscopy in gaining deeper insight into physical phenomena ranging from astronomical~\cite{penzias1965} to mesoscopic~\cite{beenakker2003,blanter2000,goennenwein2000} scales. The voltage fluctuations across a resistor in thermal equilibrium, known as the Johnson-Nyquist (JN) noise~\cite{johnson1928,nyquist1928} is attributed to the charge current fluctuations due to the random thermal motion of the charge carriers in electrical conductors. It is much less appreciated that spin current fluctuations exist in all metals since they do not interfere with most electronic processes. However, they become observable due to the spin-charge coupling in magnetic nanostructures.~\cite{foros2005,smith2001} The recently discovered spin Seebeck effect~\cite{uchida2010} is attributed to an imbalance of spin current fluctuations~\cite{xiao2010,adachi2013} caused by a thermal gradient in a ferromagnet$|$normal metal bilayer system. Spin dependent coherent transport could be detected in magnetic tunneling junctions (MTJs) via current shot noise measurements.~\cite{guerrero2006,arakawa2011} However, a direct measurement of thermal spin current noise has, to our knowledge, not been reported yet.

Here we report measurements of the voltage noise power spectral density (PSD) and resistance across a Platinum (Pt) thin film deposited on a yttrium iron garnet (YIG) layer as a function of the angle $\alpha$ between the applied magnetic field and the transport direction. These experiments are interpreted in terms of the thermal spin current noise in Pt modulated by the magnetization direction, which is transformed into charge noise by the inverse spin Hall effect (ISHE). The voltage PSD is found to obey the same angular dependence as the electric resistance, called spin Hall magnetoresistance (SMR),~\cite{nakayama2013,chen2013} consistent with the fluctuation-dissipation theorem (FDT).~\cite{landau1996} Since spin Hall effect (SHE)~\cite{kato2004,hirsch1999} is believed to be the dominant spin-charge coupling mechanism in  heavy metals films, we refer to our measurements as ``spin Hall noise" (SHN). 

The random thermal motion of the electrons in a normal metal (N) causes charge-current, but because of their spin degree of freedom, also spin-current fluctuations. The ISHE converts spin-current into charge-current (or voltage) noise. In a ferromagnetic insulator (FI)$|$N~\footnote{Here the term `ferromagnets' includes ferrimagnets such as YIG.} heterostructure, the measured voltage noise $S_V = S_V^s + S_V^{JN}$ is composed of spin induced ($S_{V}^s$) and charge ($S_{V}^{JN}$) noise. The FI modulates the conductor by selectively absorbing spin currents polarized  normal to the magnetization direction, i.e. the spin transfer torque.~\cite{maekawa2012} The implied dependence of the spin induced noise power $S_{V}^s$ on an applied magnetic field that controls the magnetization direction in the FI allows us to disentangle it from the charge noise $S_{V}^{JN}$ in the measured voltage noise $S_{V}$. A spin-charge coupling can in principle be achieved as well just at the FI$|$N interface by either the anomalous Hall effect in proximity induced ferromagnetism in N~\cite{huang2012} or a Rashba-type spin-orbit interaction.~\cite{grigoryan2014} However, there is evidence against a significant proximity effect at the YIG$|$Pt interface.~\cite{kikkawa2013,althammer2013} Furthermore, our basic result that the angular dependent thermal noise is a direct measure of spin fluctuations is model-independent.

\section{Experiments}

\begin{figure}[tb]
 \begin{center}
 \subfloat[]{\includegraphics[width=65mm]{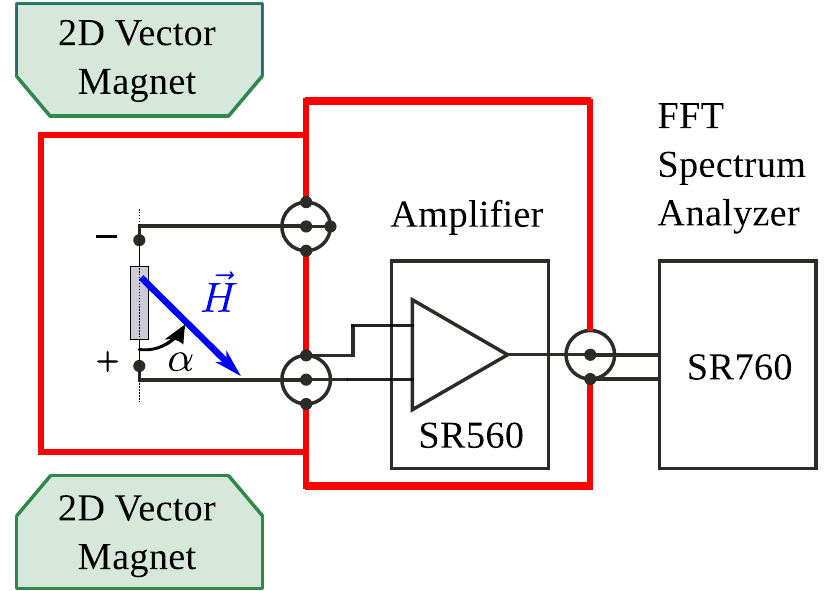}} \\
 \subfloat[]{\includegraphics[width=65mm]{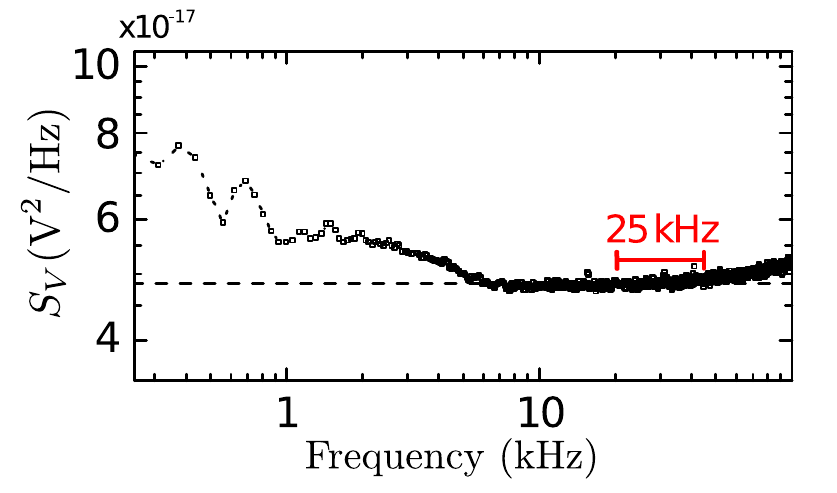}}
 \caption{(a) Schematic of the voltage power spectral density measurements. The sample (grey) is connected to a pre-amplifier and a FFT spectrum analyzer. The symbols $+$ and $-$ define the sign convention for the voltage measurements. The setup and the amplification stage are shielded by a metal box (red thick lines). The applied magnetic field (blue arrow) makes an angle $\alpha$ with the voltage measurement direction. (b) A typical noise spectrum captured using the setup described in (a). The individual data points shown in Fig. \ref{setupNsmn} (c) are averaged over the frequency window between 20 and 45 kHz. The dashed line depicts the white noise level expected from the fluctuation-dissipation theorem.} \label{circuit}
 \end{center}
\end{figure}

We first discuss the measurements of voltage noise and resistance of YIG$|$Pt bilayers. Samples were fabricated by depositing $60\,$nm of YIG ($\mathrm{Y}_{3}\mathrm{Fe}_{5}\mathrm {O}_{12}$) on a $500\,\mu$m thick, (111) oriented gadolinium gallium garnet ($\textrm{Gd}_{3} \textrm{Ga}_5 \textrm{O}_{12}$, GGG) substrate via pulsed laser deposition. A Pt film with thickness $t_N = 2.2\,$nm was then grown {\it in situ} on top of the YIG film using electron beam evaporation. Subsequently, the sample was patterned into a Hall bar mesa structure (width $w$ = 80 $\mu$m, length $l$ = 950 $\mu$m) using optical lithography and argon ion beam milling. The detailed sample preparation is described in Ref. \onlinecite{weiler2013}.

The voltage PSD was measured as sketched in Fig. \ref{circuit} (a). The voltage signal is fed into a Stanford Research fast Fourier transform spectrum analyzer (SR760) after amplification using a Stanford Research pre-amplifier (SR560). We refer to the square of the Fourier transform of a single finite duration time trace of the voltage signal as a `spectrum'. A `PSD sweep' [as in Fig. \ref{circuit} (b)] is obtained by averaging 15000 such spectra. A single average value of the white noise level is then obtained by averaging the PSD sweep data in the frequency range 20 - 45 kHz. The frequency window is so chosen in order to minimize the effects of the 1/f noise and external electromagnetic disturbances. The average of 19 such data points lead to the precision of 0.01 \% sufficient to resolve the spin Hall noise [Eq. (\ref{rhos})].~\footnote{The noise floor of our setup (output with zero voltage input i.e. short circuited amplifier input) $1.52 \times 10^{-17}\,\textrm{V}^2/$Hz is subtracted from all data points.}

\begin{figure}[tb]
 \begin{center}
  \includegraphics[width=85mm]{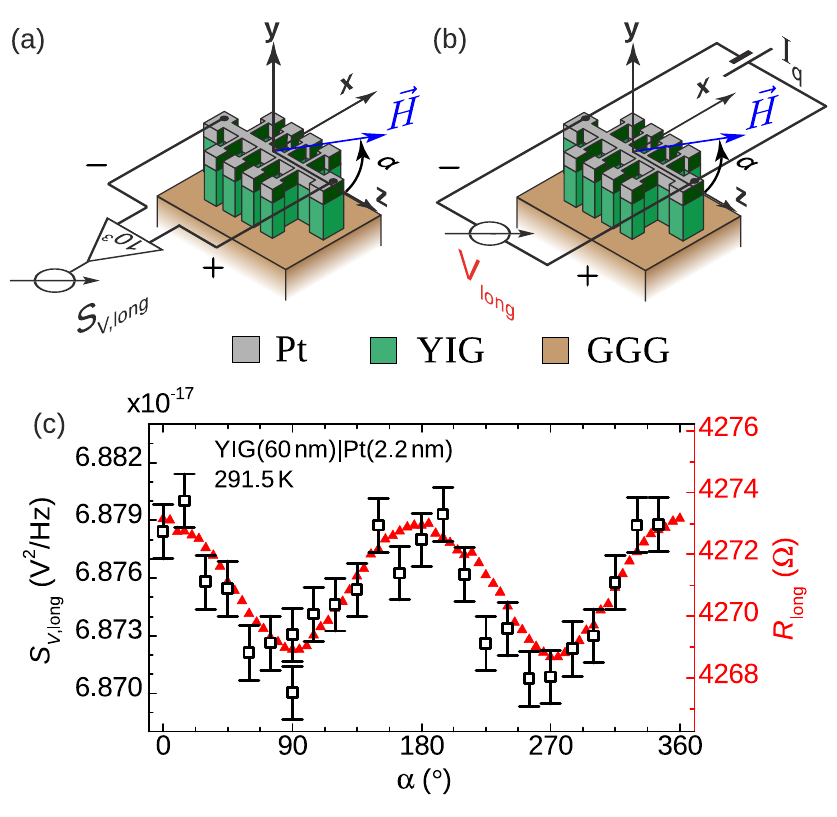}
  \caption{Sample and wire bonding assembly for measuring (a) voltage power spectral density $S_{\textrm{V,long}}$ and (b) resistance $R_{\textrm{long}}$. The applied magnetic field (blue arrow) makes an angle $\alpha$ with the voltage measurement direction ($\pmb{z}$). (c) $S_{\textrm{V,long}}$ (squares) and $R_{\textrm{long}}$ (triangles) measured as a function of $\alpha$. Both $S_{\textrm{V,long}}$ and $R_{\mathrm{long}}$ exhibit a $\cos^2 \alpha$ dependence and are related by $S_{\textrm{V,long}} = 4 k_B T R_{\textrm{long}}$ with $T = 291.5\,$K, in consistence with the fluctuation-dissipation theorem. The $\alpha$ dependent contributions to $R_{\mathrm{long}}$ and $S_{\mathrm{V,long}}$ are attributed to spin Hall effect generated spin currents and spin Hall noise, respectively.}\label{setupNsmn}
 \end{center}
\end{figure}

The measurement configuration is depicted in Fig. \ref{setupNsmn} (a). A 60 mT magnetic field applied in the $xz$-plane at an angle $\alpha$ with the $+z$-direction saturates the YIG magnetization along its direction. The voltage noise PSD $S_{\textrm{V,long}}$ of the `longitudinal' voltage $V_{\textrm{long}}$ [Fig. \ref{setupNsmn} (a)] averaged over 19 $\alpha$ sweeps is shown as white open squares in Fig. \ref{setupNsmn} (c). We also carried out conventional SMR measurement~\cite{nakayama2013} of the longitudinal resistance $R_{\textrm{long}}$ along the Hall bar ($\pmb{z}$) direction [Fig. \ref{setupNsmn} (b)] as a function of $\alpha$ for a charge current $I_{q} = 40.5\,\mu$A  along the Hall bar. $R_{\textrm{long}}$, shown as red triangles in Fig. \ref{setupNsmn} (c), exhibits the $\cos^2 \alpha$ dependence characteristic of the SMR effect.~\cite{chen2013} We find that $S_{\textrm{V,long}}$ and $R_{\textrm{long}}$ are related by $S_{\textrm{V,long}} = 4 k_B T R_{\textrm{long}}$, with T = 291.5 K (room temperature), as expected from the fluctuation-dissipation theorem. Since the $\alpha$-dependence of $R_{\mathrm{long}}$ is attributed to SHE generated spin currents,~\cite{chen2013} the anisotropic PSD must be caused by the spin Hall noise.

\section{Theory}

\begin{figure}[tp]
\begin{center}
\includegraphics[scale=0.7]{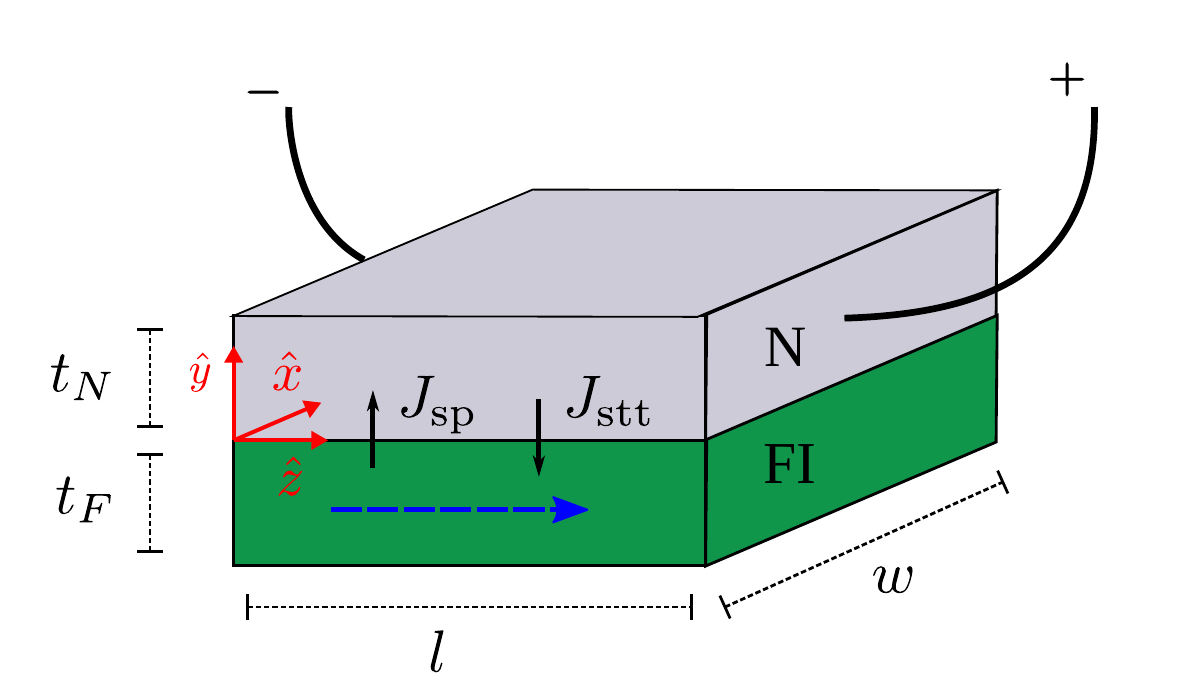}
\caption{Schematic of the normal metal (N) and magnetic insulator (FI) bilayers analyzed in the text. The blue dashed arrow indicates the equilibrium magnetization direction. The coordinate system is depicted in red. The black arrows define our sign convention for spin currents across the interface.}\label{bilayer}
\end{center}
\end{figure}

To substantiate this claim, in the following we present a statistical linear response theory for the $\alpha$ dependent noise that elucidates the role of the spin currents. We restrict the analysis to frequencies far below the ferromagnetic resonance (FMR) frequency $f_0$. We consider a bilayer of a normal metal (N) with spin Hall angle $\theta_{SH}$ deposited on a ferromagnetic insulator (FI) with its equilibrium magnetization pointing along $\hat{\pmb{z}}$ as shown in Fig. \ref{bilayer}. The magnetization dynamics in the FI is described by the Landau-Lifshitz-Gilbert (LLG) equation:
\begin{align}
 \dot{\pmb{m}} & =  - \gamma \left[ \pmb{m} \times \mu_0 (\pmb{H}_{\mathrm{eff}} + \pmb{h}_0) \right] + \alpha_0 (\pmb{m} \times \dot{\pmb{m}}),
\end{align}
where $\pmb{m} \equiv \pmb{m}(\pmb{r},t)$ is the unit vector along the magnetization direction at position $\pmb{r}$, $\gamma ( > 0)$ denotes the gyromagnetic ratio, $\pmb{h}_0$ and $\alpha_0$ the internal Langevin stochastic field~\cite{brown1963} and Gilbert damping constant, respectively. The effective magnetic field, written in terms of the magnetic free energy density $F_m$:
\begin{align}
 \mu_0 \pmb{H}_{\mathrm{eff}} = - \frac{\partial F_m}{\partial \pmb{M}} & = \mu_0 \pmb{H}_{0} + \frac{2 A}{M_s} {\nabla^2} \pmb{m}, 
\end{align}
includes Zeeman and anisotropy contributions in $\pmb{H}_0$, while the second term represents the exchange field in terms of the exchange constant $A$~\cite{kittel1949} and the saturation magnetization $M_s$. The N layer is incorporated by imposing continuity of  spin current density across the FI$|$N interface.~\cite{hoffman2013} On the FI side, the spin current density, carried by collective magnetization dynamics, is given by $- A (\pmb{m} \times  \partial_y \pmb{m})$. On the N side, the spin current density consists of $(\textrm{i})$ spin pumping ($\pmb{J}_{\textrm{sp}}$) by the thermal fluctuations of the magnetization in the ferromagnet,~\cite{tserkovnyak2002} and $(\textrm{ii})$ spin transfer torque (STT) ($\pmb{J}_{\textrm{stt}}$) generated by absorption of the thermal electronic spin current incident on the FI. The conserved net spin current density $\pmb{J}^{\textrm{s}} (=  \pmb{J}_{\textrm{sp}} - \pmb{J}_{\textrm{stt}})$ from the FI to the N is then given by:
\begin{align}\label{scont}
 \pmb{J}^{\textrm{s}}(\pmb{\varrho},t)  = & -  A (\pmb{m} \times  \partial_y \pmb{m}), \\
      = & \frac{\hbar \tilde{g}_r}{4 \pi}  (\pmb{m} \times \dot{\pmb{m}})  - M_s   (\pmb{m} \times \mu_0 \pmb{h}^\prime ) , \label{intcurrent}
\end{align}
where $\pmb{\varrho} \equiv (x,z)$ is the in-plane position vector, $\tilde{g}_r$ is the real part of the spin mixing conductance per unit area corrected for the finite thickness and/or spin relaxation length in N leading to a backflow spin current into FI.~\cite{tserkovnyak2002b} We disregard the typically small~\cite{jia2011} imaginary part of the mixing conductance for simplicity. $\pmb{h}^\prime$ represents the	 random STT with the correlation function:~\cite{foros2005,xiao2010,hoffman2013}
\begin{align}\label{correlators}
 \left\langle \mu_0 h_{i}^\prime (\pmb{\varrho},t) \mu_0 h_{j}^\prime(\pmb{\varrho}^\prime,t^\prime)   \right\rangle & =  \Sigma^\prime \delta_{ij} \delta(t-t^\prime) \delta(\pmb{\varrho}-\pmb{\varrho}^\prime),
\end{align}
where $\langle ~ \rangle$ denotes statistical averaging, $\Sigma^\prime =  \hbar \tilde{g}_r k_B T/2 \pi M_s^2 $, $(i,j) = (x,y)$, $k_B$ is the Boltzmann constant, and $T$ is the temperature of the system. 

Since the spin current flows across the interface along the out-of-plane ($y$) direction (see Fig. \ref{bilayer}), its $y$ polarized component does not contribute to the ISHE signal,~\cite{mosendz2010} while the $z$ polarized component vanishes. Hence, we focus on the $x$ component [Eq. (\ref{intcurrent})]:
\begin{eqnarray}
 J_x^\textrm{s} & = & - \frac{\hbar \tilde{g}_r}{4 \pi} \dot{m}_y + M_s \mu_0 h^\prime_{y},
\end{eqnarray}
with correlation function:
\begin{widetext}
\begin{eqnarray}
 \left\langle J_x^\textrm{s}(\pmb{\varrho},t) J_x^\textrm{s}(\pmb{\varrho}^\prime,t^\prime) \right\rangle & = &  M_s^2 \left\langle \mu_0 h_{y}^\prime (\pmb{\varrho},t) \mu_0 h_{y}^\prime(\pmb{\varrho}^\prime,t^\prime)   \right\rangle +  \left(\frac{\hbar \tilde{g}_r}{4 \pi} \right)^2 \langle \dot{m}_y(\pmb{\varrho},t) \dot{m}_y(\pmb{\varrho}^\prime,t^\prime) \rangle  \nonumber \\ 
      &  &  - \frac{\hbar \tilde{g}_r M_s \mu_0}{4 \pi} (\langle \dot{m}_y(\pmb{\varrho},t) h_y^\prime(\pmb{\varrho}^\prime,t^\prime) \rangle + \langle \dot{m}_y(\pmb{\varrho}^\prime,t^\prime) h_y^\prime(\pmb{\varrho},t) \rangle) .
\end{eqnarray}
\end{widetext}
Only the first term on the rhs of the equation above is appreciable~\cite{xiao2009} because the ac susceptibility and therefore $\dot{m}_{y}$ are negligibly small at frequencies under consideration ($f \ll f_0$). With Eq. (\ref{correlators}):
\begin{equation}
    \left\langle J_x^{\textrm{s}}(\pmb{\varrho},t) 
J_x^\textrm{s}(\pmb{\varrho}^\prime,t^\prime) \right\rangle   =  \frac{\hbar \tilde{g}_r}{2 
\pi} k_B T \delta(t-t^\prime) \delta(\pmb{\varrho}-\pmb{\varrho}^\prime). \label{spincorr}
\end{equation}
In this low frequency limit, all parameters of the ferromagnet, except for the interface spin mixing conductance, conveniently drop out.

For frequencies much smaller than the inverse spin relaxation time in N, 
the spatially resolved spin-current density is governed by
the time-independent diffusion equation $\partial^2 \pmb{\mu}_\textrm{s}/ \partial y^2 = 
\pmb{\mu}_s/ \lambda_{sd}^2$ for the spin chemical potential $\pmb{\mu}_s$	 
with the boundary conditions $\pmb{J}^\textrm{s} (= - D \partial 
\pmb{\mu}_s/\partial y) = J_x^\textrm{s}(\pmb{\varrho},t) \hat{\pmb{x}}$ at $y = 0$ 
and $\pmb{J}^\textrm{s} = 0$ at $y = t_N$:~\cite{mosendz2010}
\begin{eqnarray}\label{quasi1d}
 J_{x}^\textrm{s}(\pmb{r},t) & = & J_{x}^\textrm{s}(\pmb{\varrho},t) \ \frac{\sinh 
\left[ (t_N- y) / \lambda_{sd} \right]}{\sinh (t_N / \lambda_{sd} )}.
\end{eqnarray}
$\lambda_{sd}$ is the spin diffusion length, $D$ is the diffusion constant in N, and the spin current flows along the y direction. This quasi 1D analysis is rigorous because in-plane lateral spin diffusion does not contribute to the global emf as shown in Appendix \ref{spindiff3d}. However, locally there might be significant corrections to Eq. (\ref{quasi1d}). 

The ISHE converts the spin current density to a charge current density along the $z$ direction:
\begin{eqnarray}\label{ishejc}
 J_z^c(\pmb{r},t) & = & - \theta_{SH} \frac{2 e}{\hbar}  J_{x}^s(\pmb{r},t),
\end{eqnarray}
with $\theta_{SH}$ the spin Hall angle of N. We are interested here in the global voltage noise over the sample edges as indicated in Fig. \ref{bilayer} (see also Appendix \ref{spindiff3d}) which amounts to:
\begin{eqnarray}
 V(t) & = & - b \int J_{x}^s(\pmb{\varrho},t) d^2 \varrho, \label{volt}
\end{eqnarray}
for frequencies far below the plasma frequency, where $b \equiv (\rho \theta_{SH} e/\hbar w) \ ( 2 \lambda_{sd}/t_N) \ \tanh 
\left(t_N/2 \lambda_{sd}\right)$ and $\rho = R w t_N / l$, with $R$ the resistance of the N layer.

Employing Eqs. (\ref{spincorr}) and (\ref{volt}), we arrive at the autocorrelation:
\begin{eqnarray}
 \langle V(t) V(0) \rangle & = & b^2 \iint \langle 
J_{x}^s(\pmb{\varrho},t) J_{x}^s(\pmb{\varrho}^\prime,0) \rangle d^2\varrho \  d^2\varrho^\prime, 
\nonumber  \\
 & = & b^2 \frac{\hbar \tilde{g}_r}{2 \pi} k_B T w l \delta(t). \label{voltcorr}
\end{eqnarray}
Using the above result and the Wiener-Khintchine theorem relating the one-sided PSD $S_f(\omega)$ and the auto-correlation of a variable $f(t)$: $ S_f(\omega)  =  2 \int \langle f(t) f(0)\rangle e^{-i \omega t} dt$, PSD of the spin Hall noise reads:
\begin{eqnarray}
 S_V^s(\omega) & = & 2 \int b^2 \frac{\hbar \tilde{g}_r}{2 \pi} k_B T w l \delta(t) 
e^{-i \omega t} dt, \\
    & = & 4 k_B T \ \rho_1\frac{ l}{w t_N} = 4 k_B T R_1 , \label{sv}
\end{eqnarray}
where $R_1 = \rho_1 l /w t_N$ and
\begin{equation}\label{rhos}
 \rho_1/\rho = \left( \theta_{SH} e \ 2 \lambda_{sd} \ \tanh \left(\frac{t_N}{2 
\lambda_{sd}} \right) \right)^2 \frac{\tilde{g}_r \rho}{4 \pi \hbar t_N}.
\end{equation}
When the equilibrium magnetization direction makes an angle $\alpha$ with the voltage measurement direction ($\hat{\pmb{z}}$), the rhs of Eq. (\ref{sv}) is simply multiplied by $\cos^2 \alpha$,~\cite{xiao2009} because only the $z$ projection ($\cos \alpha$) of the fluctuating ISHE current [Eq. (\ref{ishejc})] contributes to the voltage fluctuations. The thermal JN noise ($S_V^{JN} = 4 k_B T R$) can be added to obtain the total voltage noise:
\begin{equation}\label{mainth}
S_{V}(\omega) = 4 k_B T (R + R_1 \cos^2 \alpha).
\end{equation} 
Our direct derivation of the PSD [Eqs. (\ref{rhos}) and (\ref{mainth})] is consistent with the FDT combined with the angle dependent resistance.~\cite{nakayama2013,chen2013} Thus, the present analysis can be considered an alternative derivation of the SMR effect.


\section{Conclusion}
In summary, we report, to the best of our knowledge, the first observation of what we call spin Hall noise. The magnetization direction-dependent voltage noise and resistance measured in a YIG$|$Pt bilayer obey the fluctuation-dissipation theorem (FDT) confirming that spin Hall current based physics of the magnetoresistance (SMR)~\cite{nakayama2013,chen2013} implies the presence of the spin Hall noise. A theoretical description for the latter in terms of spin current fluctuations gives insight into the non-trivial nature of entanglement of the spin contribution with the magnetization dynamics. In light of the FDT, observation of spin current fluctuations emphasizes the dissipative nature of pure spin currents. The experimental resolution of the spin current noise demonstrated here paves the way for advanced noise spectroscopy studies, such as (non-equilibrium) resistance~\cite{foros2007} and spin pumping shot noise.


\section*{Acknowledgments}
AK thanks J. Xiao, Y. T. Chen, M. Wimmer, and S. Sharma for fruitful discussions. Financial support from the DFG via SPP 1538 ``Spin Caloric Transport'', Project Nos. GO 944/4-1 and BA 2954/1, Grant-in-Aid for Scientific Research (Kakenhi) 25247056/25220910/26103006, the Dutch FOM Foundation and EC Project ``InSpin'' is gratefully acknowledged.

\appendix


\section{Spin diffusion in 3D}\label{spindiff3d}
Here, we solve the spin diffusion equation in the normal metal (N) and calculate the spin current correlators required for evaluating the voltage noise power spectral density (PSD). We show that a three dimensional analysis yields the same result as the quasi-one dimensional model [Eq. (\ref{quasi1d})]. The notations and the coordinate system are defined in Fig. \ref{bilayer}.

Since we are interested in the $x$ polarized component of the spin current ($\pmb{J}_x^s(\pmb{r},t) = - D \pmb{\nabla} \mu_x^s$) at frequencies much smaller than the inverse spin flip rate, we have to solve the time-independent spin-diffusion equation:
\begin{align}\label{diffeq1}
 \nabla^2 \mu_x^s & = \frac{\mu_x^s}{\lambda_{sd}^2},
\end{align}
with the boundary conditions $J_{x}^s  \hat{\pmb{y}} =  - D \partial_y \mu_x^s= J_{x}^s(\pmb{\varrho},t)$ [see Eq. (\ref{intcurrent})] at $y = 0$ and $J_{x}^s  \hat{\pmb{y}} = 0$ at $y = t_N$, where $J_{x}^s  \hat{\pmb{y}}$ denotes the $x$ polarized spin current flowing along the $y$ direction. This equation is valid for frequencies much smaller than the spin flip rate ($\sim$ THz in Pt). Physically, all time dependence comes from the boundary conditions to which the spin accumulation reacts instantaneously. The general solution for a translationally invariant planar system reads:~\cite{morse1953,mosendz2010}
\begin{align}
 \mu_x^s & = \sum_{\pmb{k}} \frac{\tilde{\mu}_x^s(\pmb{k})}{\mathcal{A}} e^{i \pmb{k}\cdot \pmb{\varrho}  } \frac{\cosh [c_y^{\pmb{k}} (t_N - y)]}{\sinh (c_y^{\pmb{k}}	 t_N)}, 
\end{align}
where $\mathcal{A}$ is the interface area, $\pmb{k}$ an in-plane wave vector, $c_y^{\pmb{k}} = \sqrt{ 1/ \lambda_{sd}^2 + k_x^2 + k_z^2}$ and spin current:
\begin{align}\label{res1}
 J_x^s(\pmb{r},t) & = \sum_{\pmb{k}} \frac{\tilde{J}_x^s(\pmb{k},t)}{\mathcal{A}} e^{i \pmb{k} \cdot \pmb{\varrho} } f(\pmb{k},y),
\end{align}
with
\begin{equation}
 f(\pmb{k},y) = \frac{\sinh [c_y^{\pmb{k}} (t_N - y)]}{\sinh (c_y^{\pmb{k}} t_N)}.
\end{equation}

The voltage auto-correlation and PSD are governed by the integral over the metal film:
\begin{align}
 g(t) & = \iint \langle J_x^s(\pmb{r},t) J_x^s(\pmb{r}^\prime,0) \rangle d^3 r d^3 r^\prime \\ 
   & = \iint F(\pmb{r},\pmb{r}^\prime,t) d^3 r d^3 r^\prime, \label{gt}
\end{align}
where $\langle \cdot \rangle$ denotes statistical averaging. With Eq. (\ref{res1}):
\begin{widetext}
\begin{align}\label{F1}
 F(\pmb{r},\pmb{r}^\prime,t) & = \frac{1}{\mathcal{A}^2} \sum_{\pmb{k},\pmb{k}^\prime} \left\langle \tilde{J}_x^s(\pmb{k},t) \tilde{J}_x^s(\pmb{k}^\prime,0) \right\rangle e^{i \left( \pmb{k} \cdot \pmb{\varrho} +  \pmb{k}^\prime \cdot \pmb{\varrho}^\prime \right)} f(\pmb{k},y) f(\pmb{k}^\prime,y^\prime).
\end{align}
Due to the boundary condition at $y = 0$, $\tilde{J}_x^s(\pmb{k},t)$ is the Fourier transform of $J_{x}^s(\pmb{\varrho},t)$, whence, employing Eq. (\ref{spincorr}),
\begin{align}
 \left\langle \tilde{J}_x^s(\pmb{k},t) \tilde{J}_x^s(\pmb{k}^\prime,0) \right\rangle & =  \iint \left\langle \tilde{J}_x^s(\pmb{\varrho},t) \tilde{J}_x^s(\pmb{\varrho}^\prime,0) \right\rangle e^{-i  \left( \pmb{k} \cdot \pmb{\varrho} +  \pmb{k}^\prime \cdot \pmb{\varrho}^\prime \right)} d^2 \varrho d^2 \varrho^\prime, \\
 & = \frac{\hbar \tilde{g}_r}{2 \pi} k_B T \ \mathcal{A} \ \delta_{\pmb{k},-\pmb{k}^\prime} \ \delta(t),
\end{align}
and
\begin{align}
 F(\pmb{r},\pmb{r}^\prime,t) & = \frac{1}{\mathcal{A}} \sum_{\pmb{k}} \frac{\hbar \tilde{g}_r}{2 \pi} k_B T \delta(t) e^{i \pmb{k} \cdot (\pmb{\varrho} - \pmb{\varrho}^\prime)} f(\pmb{k},y) f(\pmb{k}^\prime,y^\prime).
\end{align}
Therefore,
\begin{align}
 g(t) & = \iint \frac{1}{\mathcal{A}} \sum_{\pmb{k}} \frac{\hbar \tilde{g}_r}{2 \pi} k_B T \delta(t) e^{i \pmb{k} \cdot (\pmb{\varrho} - \pmb{\varrho}^\prime)} f(\pmb{k},y) f(\pmb{k}^\prime,y^\prime) d^3 r d^3 r^\prime, \\
   & = \frac{\hbar \tilde{g}_r}{2 \pi} k_B T \mathcal{A} \delta(t) \left( \int_0^{t_N}  \frac{\sinh \left( \frac{t_N - y}{\lambda_{sd}} \right) }{\sinh \left( \frac{t_N}{\lambda_{sd}} \right)} dy \right)^2, \\
   & = \frac{\hbar \tilde{g}_r}{2 \pi} k_B T \mathcal{A} \lambda_{sd}^2 \tanh^2 \left( \frac{t_N}{2 \lambda_{sd}} \right) \ \delta(t),
\end{align}
\end{widetext}
which agrees with Eq. (\ref{voltcorr}). The volume integral of the emf that amounts to the total voltage across N corresponds to the $k = 0$ component of the in-plane variations thereby reducing the 3D to an effectively 1D problem.

\bibliography{SHN}

\end{document}